\documentclass[%
 reprint,
 amsmath,amssymb,
 aps,
]{revtex4-2}
\usepackage{arydshln}
\usepackage{multirow}

\usepackage{graphicx}% Include figure files

\usepackage{dcolumn}% Align table columns on decimal point
\usepackage{bm}% bold math

\usepackage[colorlinks,linkcolor=blue,citecolor=blue,urlcolor=blue]{hyperref}

\begin{document}

\preprint{}

\title{Thermodynamic origin of the phonon  Hall effect   in a honeycomb antiferromagnet}

\author{Qingkai Meng$^{1}$, Xiaokang Li$^{1,*}$, Jie Liu$^{1}$, Lingxiao Zhao$^{2}$, Chao Dong$^{1}$, Zengwei Zhu$^{1,*}$, Liang Li$^{1,*}$ and Kamran Behnia$^{3,*}$}

\affiliation{
(1) Wuhan National High Magnetic Field Center, School of Physics, Huazhong University of Science and Technology, 430074 Wuhan, China\\ 
(2) Quantum Science Center of Guangdong-HongKong-Macao Greater Bay Area, 518045 Shenzhen, China\\ 
(3) Laboratoire de Physique et d'\'Etude des Mat\'eriaux \\ (ESPCI - CNRS - Sorbonne Universit\'e), PSL Research University, 75005 Paris, France
}
\date{\today}

\begin{abstract}
The underlying mechanism of the thermal Hall effect (THE) generated by phonons in a variety of insulators is yet to be identified. Here, we report on a sizeable thermal Hall conductivity in NiPS$_3$, a van der Waals stack of honeycomb layers with a zigzag antiferromagnetic order below $T_N$ = 155 K. The longitudinal ($\kappa_{aa}$) and the transverse ($\kappa_{ab}$) thermal conductivities peak at the same temperature and the thermal Hall angle, at this peak, respects a previously identified bound. The amplitude of $\kappa_{ab}$ is extremely sensitive to the amplitude of magnetization along the $b$-axis, in contrast to the phonon mean free path, which is not at all. We show that the magnon and acoustic phonon bands cross each other along the $b^\ast$ orientation in the momentum space. The relevance of a thermodynamic property, combined with the irrelevance of the mean free path, points to an intrinsic origin.

%The exponential temperature dependence of $\kappa_{ab}$ above its peak reveals an energy scale on the order of magnitude of the gap expected to be opened by magnon-phonon hybridization. 

\end{abstract}

\maketitle
The thermal analog of electrical Hall effect is dubbed thermal Hall effect (THE). It refers to a transverse temperature difference generated by a longitudinal heat current in presence of a perpendicular magnetic field. For a long time, phonons, lacking charge and spin, were thought incapable of contributing to the THE~\cite{Perkins2022}. However, this assumption was proven wrong by the discovery of a THE in a paramagnetic insulator (Tb$_3$Ga$_3$O$_{12}$) in 2005~\cite{Strohm2005}. Numerous experiments~\cite{Hirsch2015,Ideue2017,Sugii2017,Hentrich2019,Grissonnanche2019,Grissonnanche2020,Boulanger2020,Li2020,Akazawa2020,Yamashita2020,Sim2021,Chen2022,Uehara2022,Jiang2022,Bruin2022,Li2023,Gillig2023,sharma2024phonon} have reported on a detectable THE in a wide variety of insulators, including three non-magnetic solids~\cite{Li2020,Li2023,sharma2024phonon}. These experimental observations motivated many theoretical studies on how the lattice vibrations (phonons) can couple to the magnetic field~\cite{Sheng2006, Kagan2008, Zhang2010, Qin2012,Agarwalla2011,Chen2020,Yang2020,Sun2022,Perkins2022,Flebus2022,Mangeolle2022,Guo2022}. 
%The main proposals can be broadly classified as extrinsic, invoking scattering of phonons, or intrinsic, referring to dispersion of phonons. 

\begin{figure}[ht]
\centering
\includegraphics[width=1\linewidth]{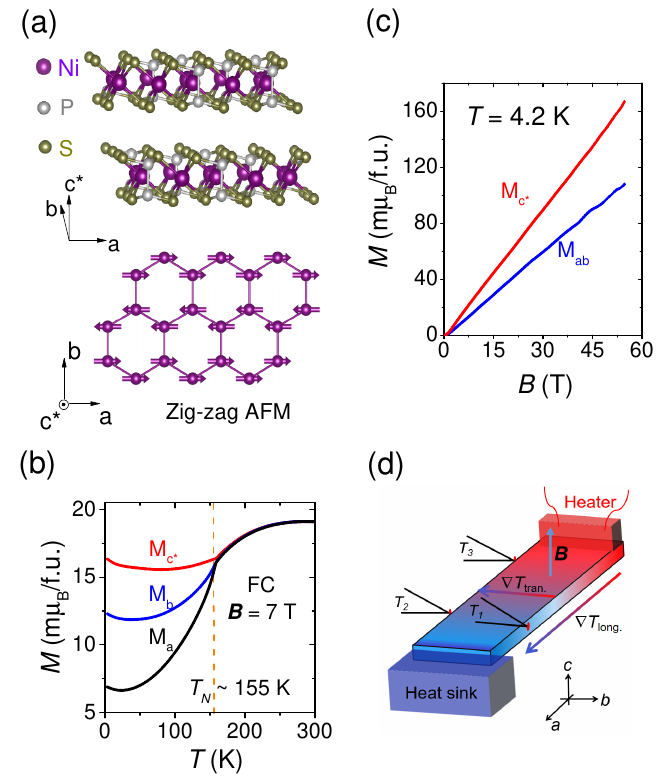} 
\caption{\textbf{Lattice structure, magnetization and thermal transport setup.} (a) Lattice structure of NPS$_3$. The side view (top) shows the layered structure. In each layer, Ni atoms (in purple) form a honeycomb lattice. The magnetic moments are arranged as zigzag chains ferromagnetically along $a-$axis and antiferromagnetically along $b-$axis below the  N\'eel temperature~\cite{Kang2020,Wildes2022}. (b) Temperature dependence of magnetization at 7 T with three different orientations. The curves deviate from each other at 155 K, suggesting the antiferromagnetic phase transition, consistent with the previous literature~\cite{Wildes2015}. (c) High field magnetization up to 55 T at 4.2 K. Absence of transition indicates a robust antiferromagnetic exchange interaction. (d) Setup for simultaneous measurements of longitudinal ($\nabla T_a = -(T_3-T_2)/l$) and transverse ($\nabla T_b = -(T_2-T_1)/w$) thermal gradients.
}
\label{fig:config}
\end{figure}

NiPS$_3$ is a layered antiferromagnet, with Ni atoms forming a honeycomb lattice (See Figure \ref{fig:config}a) stacked on top of each other  and coupled by van der Waals interaction~\cite{Wildes2015,Kim2018}. Its zigzag antiferromagnetic order can be described by a Hamiltonian with ferromagnetic exchange between the first and the third neighbors and a much weaker antiferromagnetic exchange between the second neighbors~\cite{Lanccon2018,Wildes2022}. It has an energy gap of 1.8 eV \cite{Kim2018} and belongs to the the family of  $M$P$X_3$ ($M =$ Fe, Mn, Ni, and $X =$ S, Se) van der Waals magnets. Exfoliation  of the bulk has shown the magnetic order of NiPS$_3$ is fully and abruptly suppressed in the monolayer limit. This implies the importance of the interlayer coupling for the magnetic order~\cite{kim2019}. The discovery of spin-orbit-entangled excitons associated with a Zhang-Rice singlet-triplet transition~\cite{Kang2020}, has amplified the interest in this family as a platform for studying many-body coherence and coupling~\cite{Liu2021,Klaproth2023,Cui2023,Luo2023}. Recently, infrared and Raman spectroscopy detected hybridization between magnon and optical phonon bands in FePSe$_3$ \cite{Liu2021,Cui2023,Luo2023}.

Here, we present a study of thermal transport in NiPS$_3$ crystals. The thermal Hall conductivity, $\kappa_{ab}$,  peaks to a value large as -1.63 W/Km at 25 K and 14 T, close to the largest ever recorded~\cite{Li2023}. The longitudinal ($\kappa_{aa}$) and transverse ($\kappa_{ab}$) thermal conductivities peak at the same temperature. The thermal Hall angle ($\frac{1}{B}\frac{\kappa_{ab}}{\kappa_{aa}}$) remains within a narrow range of $10^{-4}$-$10^{-3}$ T$^{-1}$. These features are shared by other insulators reported to display a THE~\cite{Li2020,Chen2022,Li2023}, including  non-magnetic ones.  The most revealing observation is a striking contrast between the sensitivity of the transverse and the insensitivity of the longitudinal  thermal conductivity to the in-plane anisotropy of the magnetization, and the angle -dependence of the free energy. Scrutinizing five different samples, we find that a small variation in the  amplitude of magnetization along $b$-axis does not affect $\kappa_{aa}$ and $\kappa_{bb}$ but enhances $\kappa_{ab}$ by an order of magnitude. This means that the amplitude of $\kappa_{ab}$ correlates with a thermodynamic property, the angle dependence of the magnetic free energy, but not with the phonon mean free path. An intrinsic origin is also backed by comparing the  computed phonon dispersion with the experimentally extracted magnon spectrum. We show that along $b^\ast$ in the reciprocal space, an acoustic phonon band and a magnon band in NiPS$_3$ cross at an energy of 3.7 meV. Interestingly, the temperature dependent of THE above its peak yields an energy scale tantalizingly close to the gap expected in the case of magnon-phonon hybridization. 

 NiPS$_3$ single crystals used in this work were grown by the chemical vapor transport method using I$_2$ as the transport agent (see the supplement for more details~\cite{SM}). %It has a monoclinic structure with the magnetic nickel atoms forming layered honeycomb lattice, and the magnetic moments arranged ferromagnetically along the zigzag chains but antiferromagnetically between adjacent chains, as shown in Figure~\ref{fig:config}a. 
 The temperature dependence of the magnetization can be seen in Figure~\ref{fig:config}b.  The N\'eel temperature is 155 K and the easy axis along  $a$, consistent with the previous report~\cite{Wildes2015,Kang2020,Wildes2022}. As seen in Figure~\ref{fig:config}c, the field dependence of magnetization up to 55 T (at 4.2 K) does not show any transition.  Given the magnetic moment of Ni atoms (1.05 $\mu_B$) ~\cite{Wildes2015}), at this field, the Zeeman energy is 3.2 meV. The persistence of the magnetic order is consistent with the largest exchange parameter quantified by inelastic neutron scattering ($J_3$ = 6.9 meV)~\cite{Lanccon2018,Wildes2022}. 

We simultaneously measured longitudinal and transverse thermal transport  using the one-heater-three-thermocouples method, shown in Figure~\ref{fig:config}d (see the supplement for more details~\cite{SM}). Figure~\ref{fig:ThermalHall}a displays the temperature dependence of $\kappa_{aa}$ at 0 T and 14 T, which superpose on top of each other. No strong anomaly is detectable at N\'eel temperature, confirming that phonons as the exclusive carriers of heat. $\kappa_{aa}(T)$ peaks to 274 W/Km, larger than most magnetic insulators, slightly lower than in Cu$_3$TeO$_6$~\cite{Chen2022}, and well below what is observed for phonons in non-magnetic insulators like black phosphorus \cite{Li2023} (or diamond, where the peak exceeds $10^4$ W/Km \cite{Onn1992}), confirming the marginal role of magnons in carrying heat.

\begin{figure*}[ht]
\centering
\includegraphics[width=0.9\linewidth]{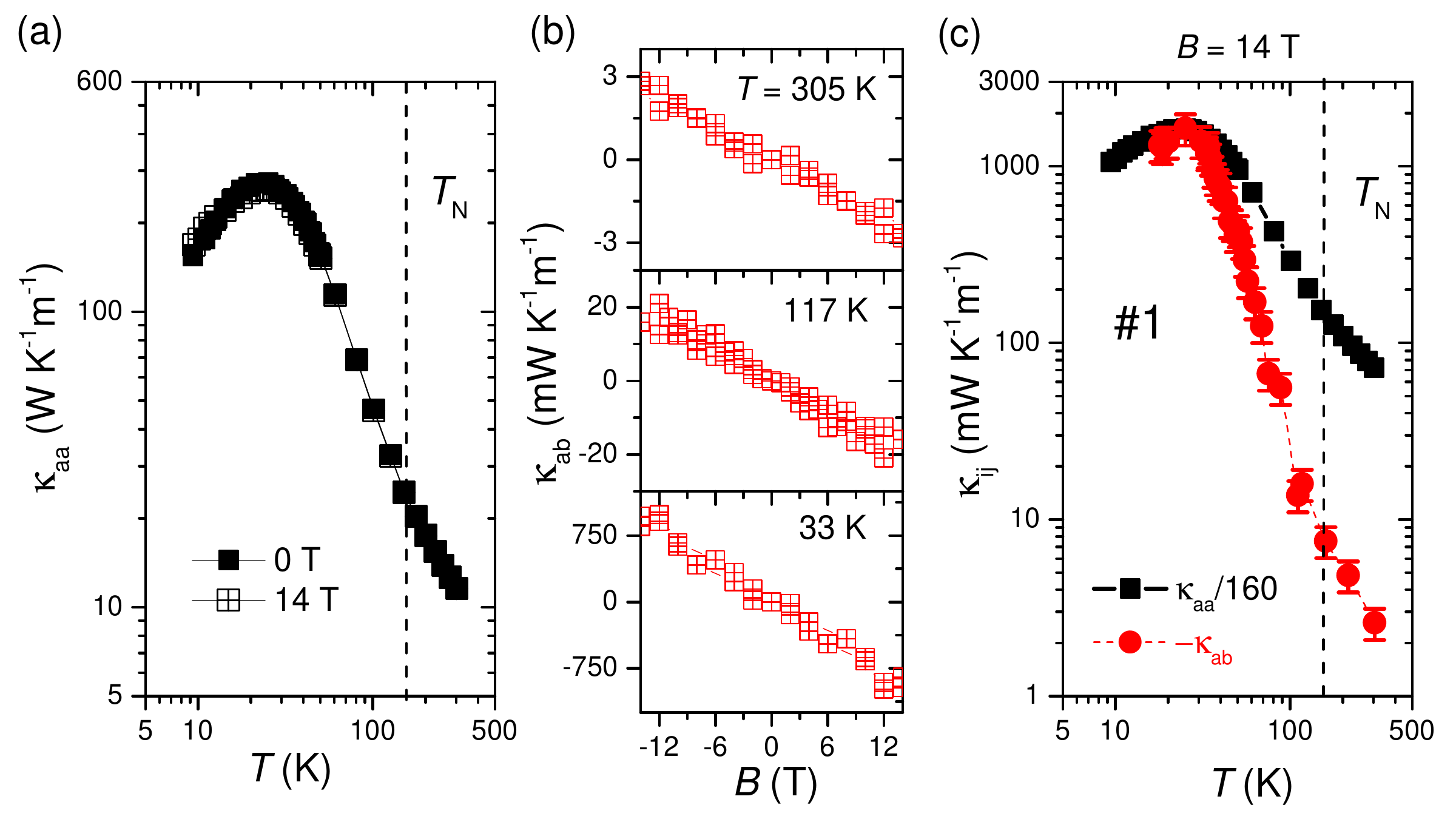} 
\caption{\textbf{Longitudinal and transverse thermal conductivities of NiPS$_3$.} (a) Temperature dependence of longitudinal thermal conductivity ($\kappa_{aa}$) under a magnetic field of 0 T and 14 T respectively. Two curves coincide well each other in the whole temperature region with a peak around 25 K. (b) Field dependence of transverse thermal conductivity ($\kappa_{ab}$) at different temperatures with the field sweeping to 14 T. In thermal Hall measurements, the heat current is along $a$ ($x$) axis, the transverse temperature gradient is along $b$ ($y$) axis. (c) Comparison of the temperature dependent $\kappa_{aa}$ (divided by -160) and $\kappa_{ab}$. 
}
\label{fig:ThermalHall}
\end{figure*}

Figure~\ref{fig:ThermalHall}b shows the field dependence of the thermal Hall conductivity $\kappa_{ab}$ at three typical temperatures and up to 14 T. The field dependence is always linear, but the amplitude increases drastically from -2.6 mW/Km at 305 K to -910 mW/Km at 33 K. This is much larger than the enhancement of $\kappa_{aa}$, implying the thermal Hall angle, the ratio of the two, also increases when cooling. 

Figure~\ref{fig:ThermalHall}c displays the temperature dependence of $\kappa_{aa}$ (divided by a factor of -160) and $\kappa_{ab}$ at 14 T. $\kappa_{ab}$ does not show any jump at $T_N$ and peaks, at 25 K and 14 T, to  attain a remarkably large peak of -1636 mW/Km. As observed in other cases \cite{Li2020,Li2023}, $\kappa_{aa}(T)$ and $\kappa_{ab}(T)$ peak at the same temperature.  Moreover, the thermal Hall angle, defined as the ratio of $\kappa_{ab}$ to $\kappa_{aa}$ divided by the magnetic field, peaks to $4\cdot10^{-4}$ T$^{-1}$ at this temperature. This remains within a narrow range of $10^{-4}$-$10^{-3}$ T$^{-1}$, where all reported maximal thermal Hall angles lie \cite{Li2023} (see the supplement for more details\cite{SM}).

The phonon mean free path at peak temperature in these materials varies from 10 nm to 100 $\mu$m. Since the thermal length scale extracted from the experimentally measured maximum Hall angle, $\lambda_{tha}=\sqrt{\frac{\hbar}{e B} \cdot \kappa_{ab}/\kappa_{aa}}$ shows little variation, it may be linked to two other relevant length scales (the interatomic distance  and the phonon wavelength at the peak temperature) which in contrast to the phonon mean free path, do not vary significantly across these insulating solids.

\begin{figure*}[ht]
\centering
\includegraphics[width=1\linewidth]{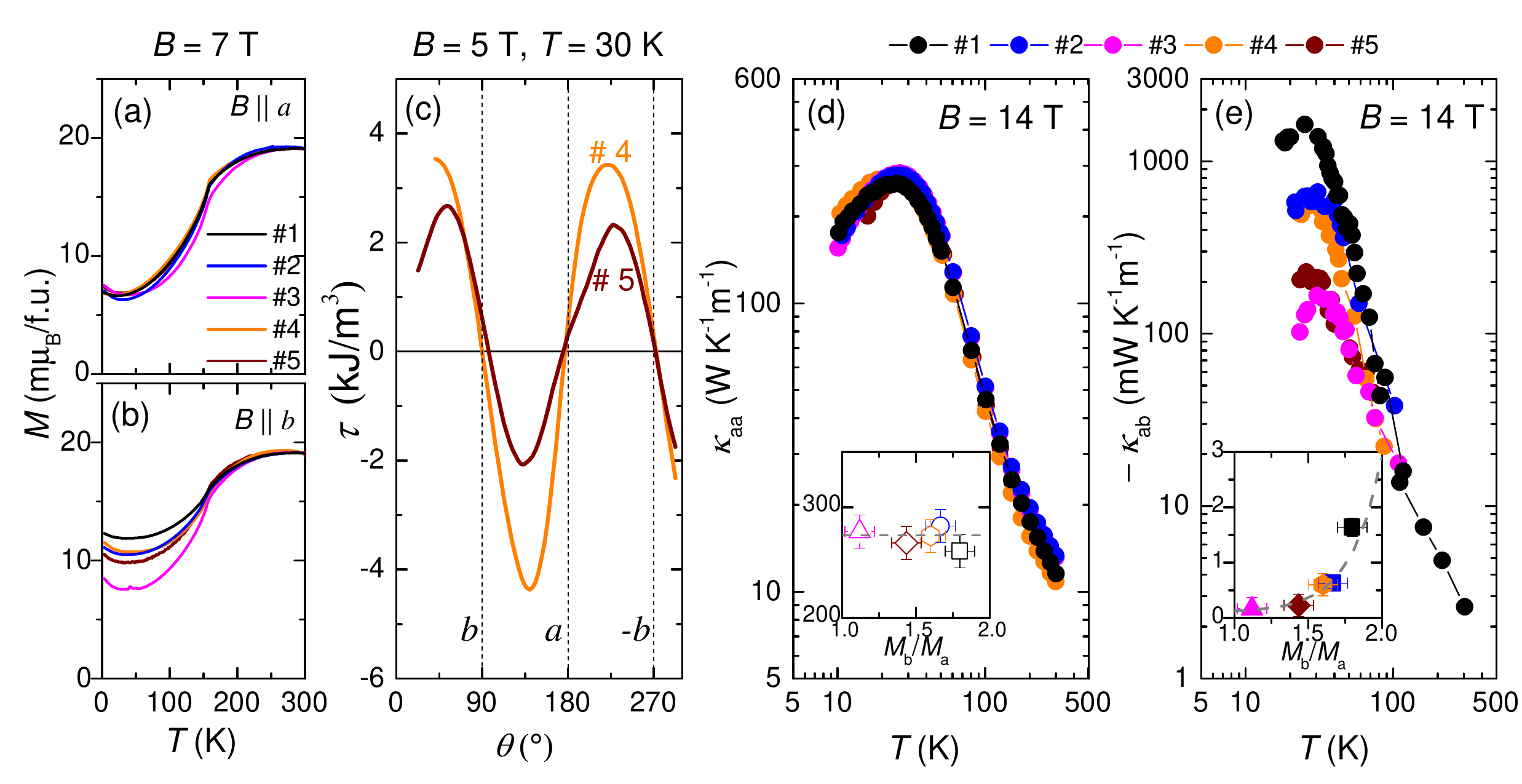} 
\caption{\textbf{Sample dependence of thermal transport, in-plane magnetization and torque in NiPS$_3$.} In-plane magnetization for $B//a$ (a) and $B//b$ (b) in five different samples (\#1 to \#5). There is no detectable change in $M_a$, but $M_b$ changes by about 30\%. (c) The angle dependence of torque at 30 K and 5 T in two samples (\#4 and \#5). The sample with a higher $M_b$ (\#4) has a larger torque response. (d) Temperature dependence of longitudinal thermal conductivity, $\kappa_{aa}$, in different samples. The curves overlap. The inset plots the peak $\kappa_{aa}$ as a function of the in-plane magnetic anisotropy ($M_b/M_a$) in the five samples. (e)  Temperature dependence of the thermal Hall conductivity, $\kappa_{ab}$ in the five samples. Contrary to the $\kappa_{aa}$, the $\kappa_{ab}$ is strongly  sample-dependent. Inset shows the peak of $\kappa_{ab}$ as a function of $M_b/M_a$. There is a clear correlation.}
\label{fig:sampledependent}
\end{figure*}

\begin{figure*}[ht]
\centering
\includegraphics[width=1\linewidth]{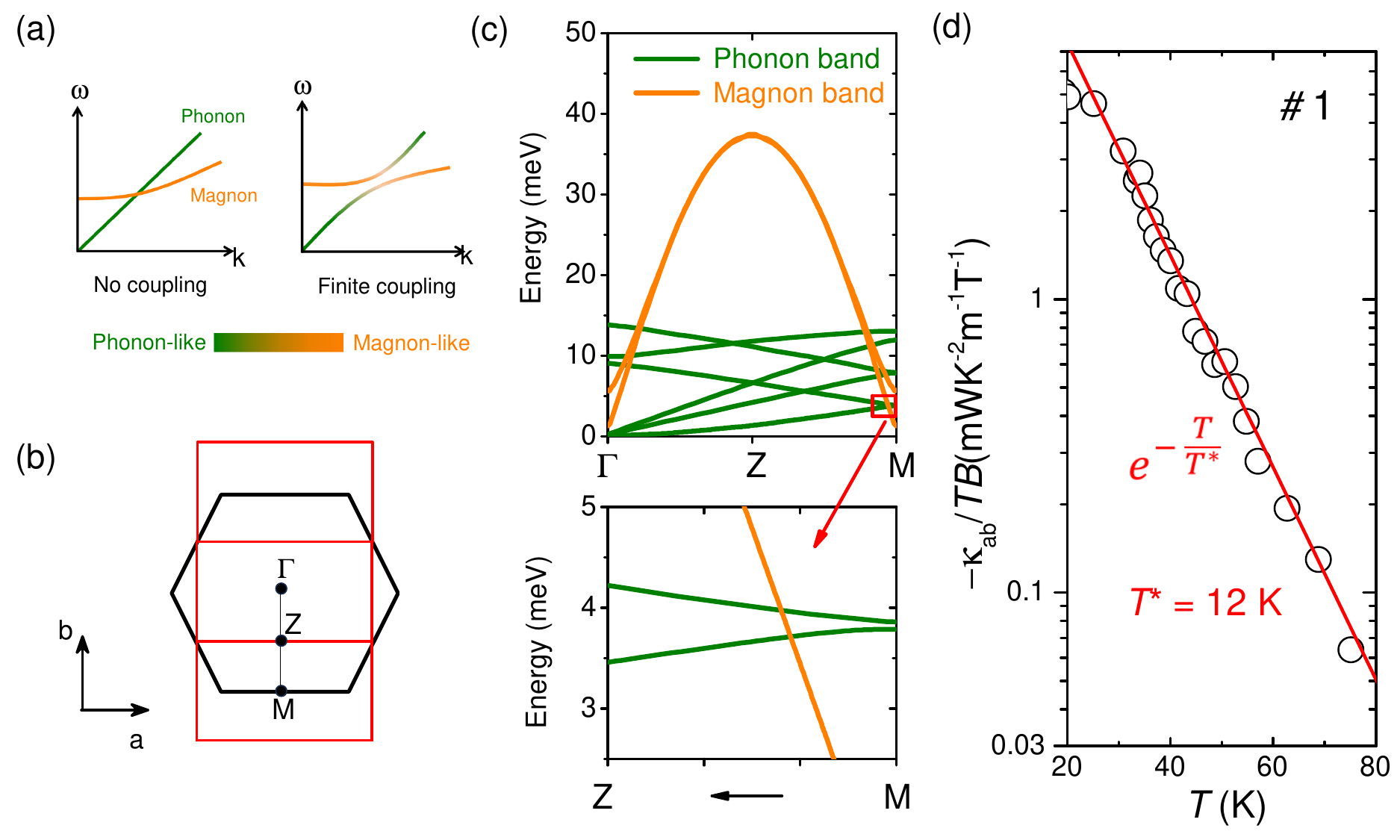} 
\caption{\textbf{Magnon-phonon hybridization and the emergence of a $\approx$ 3 meV energy scale.}  (a) Schematic diagram of the magnon-phonon band hybridization (adapted from ref.~\cite{Nomura2019}). 
 (b) Brillouin zone boundaries for magnetic (in red) and crystallographic (in black) Brillouin zones \cite{Wildes2023}. Note the M-point, which is at the boundary of a crystallographic Brillouin zone and at the center of a magnetic one.  (c) The calculated phonon spectrum (in green)~\cite{hashemi2017} and the experimentally derived magnon spectrum~\cite{Wildes2022,Scheie2023} of NiPS$_{3}$. Note that the magnon spectrum is identical along $\Gamma$-Z and along M-Z. This leads to a magnon-phonon crossing near the M-point. As seen in the lower panel, which is a zoom on the lower crossing points occurring near 4 meV.  (d) A semi-log plot of $\kappa_{ab}/TB$ as a function of temperature in the sample \#1. $\kappa_{ab}/TB$ peaks at $\approx$ 30 K. Above this temperature, it follows $exp(-T/T^\star)$, with $T^\star=12$ K.
 }
\label{fig:mechanism}
\end{figure*}

Proposed theoretical mechanisms of phonon THE~\cite{Sheng2006, Kagan2008, Zhang2010, Qin2012,Agarwalla2011,Chen2020,Yang2020,Sun2022,Perkins2022,Flebus2022,Mangeolle2022,Guo2022} can be broadly divided into extrinsic and intrinsic. Most extrinsic scenarios invoke the phonon mean free path. For example, according to a recent theoretical proposal~\cite{Sun2022}, longitudinal and transverse thermal conductivities can be simply expressed as: $\kappa_{aa} \sim C \upsilon^2 \tau \propto \ell $ and $\kappa_{ab} \sim C \upsilon^2 \tau^2 \tau_0^{-1} \propto \ell^2$. Here, $C$, $\upsilon$, $\tau$, $\ell$ ($=\upsilon \tau$) and $\tau_0^{-1}$ are  respectively the phonon specific heat, the sound speed, the average scattering time, the mean free path  and the time-reversal odd skew scattering rate. The two equations imply  $\kappa_{aa} \propto \ell$, $\kappa_{ab} \propto \ell^2$, and $\kappa_{ab}/\kappa_{aa}\propto \ell$, in contrast to experiment, which finds that $\kappa_{ab}/\kappa_{aa}$ does not depend on $\ell$.
 
Studying five crystals of NiPS$_3$, we found an identical longitudinal thermal conductivity and a sample-dependent thermal Hall response  correlating with the anisotropy of magnetization and the presence of twins. The results are shown in Figure~\ref{fig:sampledependent}. The temperature dependence of magnetization $M$ at 7 T in five  different samples is shown in Figure~\ref{fig:sampledependent}a,b. While $M_a$ is almost identical, $M_b$ displays a sample variation by approximately $\approx$ 30\% between the most and the least anisotropic samples. Our data is consistent with reported by two other groups (see the supplement for more details~\cite{SM}).  The sample dependence of magnetization can be understood, given the presence of twins in  this family \cite{murayama2016,Scheie2023}. Twinned domains share a common $c^*$-axis, but are rotated by 120 degrees off each other (see the supplement for more details~\cite{SM}).  The difference between magnetization along the $a$-axis and along the $b$-axis is attenuated by the presence of minority domains in which the spin lattice and the crystal lattice are rotated by 120 degrees. Sample $\#1$, with the highest magnetic anisotropy is the closest to a perfect untwinned sample and sample $\#3$ is the one where there is almost the same population of three possible crystalline domains. Our torque magnetometry data on two samples (Figure~\ref{fig:sampledependent}c) confirm this picture.  The angle dependence of the magnetic free energy shows a two-fold oscillation vanishing at the two high symmetry axes. Its amplitude is larger in the sample with the larger $\frac{M_b}{M_a}$ (sample \#4).  Figure~\ref{fig:sampledependent}d, e  shows the temperature dependence of $\kappa_{aa}$ and $\kappa_{ab}$ in five different samples. $\kappa_{aa}$ curves all fall on top of each other. This is also the case of the $\kappa_{bb}$ and the specific heat (see the supplement for more details~\cite{SM}). Despite an identical phonon mean free path, the transverse response $\kappa_{ab}$ is very different in these samples. The peak amplitude is $\sim 10$ times larger in sample \#1 than in sample \#3. 

The correlation between the amplitude  of the thermal Hall conductivity and a component of magnetization is reminiscent of the  Streda formula linking a Hall response to a thermodynamic property~\cite{Streda_1982,Zhang2020}. In sharp contrast, the longitudinal thermal conductivity and phonon mean free path are isotropic and do not  correlate with $\kappa_{ab}$ or with $M_b/M_a$.
 
NiPS$_3$ is not the first case of a THE with variable amplitude. Replacing a tiny fraction of Sr atoms with Ca strongly damps $\kappa_{xy}$ in SrTiO$_3$ \cite{Jiang2022}.  In Sr$_2$IrO$_4$ \cite{ataei2024phonon}, replacing Ir with Rh first amplifies $\kappa_{xy}$ before suppressing it when the magnetic order is destroyed. In both these cases, however, the large change in  the amplitude of the thermal Hall angle is accompanied by a change in the longitudinal thermal conductivity and a modification of the ground state. In the present case, we observed a tenfold variation of the thermal Hall response with no change in the longitudinal transport. The change in the magnetic response implies that the change is driven by the re-alignment between the spin and the crystal lattices.

Comparing NiPS$_3$ with $\alpha$-RuCl$_3$ is also instructive. Both are  van der Waals layers of a honeycomb lattice with a zigzag antiferromagnetic order at zero temperature. However, in the latter, in contrast to the former, one suspects the presence of a sizeable Kitaev term. In this context, it is striking that the longitudinal and the transverse thermal conductivity are both a hundred times larger in NiPS$_3$ than  in $\alpha$-RuCl$_3$ (see the supplement for more details~\cite{SM}). This observation has profound implications. First, the replacement of well-defined magnons in NiPS$_3$ by  a continuum of incoherent magnetic excitations in $\alpha$-RuCl$_3$ \cite{Winter2017} leads to a severe damping of the phonon mean free path. Nevertheless, the thermal Hall angle of the two solids is roughly equal. What sets this angle neither scales with the phonon mean free path nor depends much on the magnetic excitations coupling with phonons.

Theoretically, degenerate chiral phonons are known to reside at the center of the Brillouin zone in a honeycomb lattice \cite{Zhang2015}. The magnetic order in NiPS$_3$, which breaks inversion center, lifts this degeneracy.  Recently, Raman scattering and infrared conductivity have reported on the hybridization between magnons and optical phonons in a sister compound of NiPS$_3$, namely FePSe$_3$~\cite{Liu2021,Cui2023,Luo2023}. However, heat is carried by acoustic phonons. Coupling between the latter and magnons has been suggested in other cases~\cite{Nomura2019,Vzivkovic2019,Zhang2019}. Suppose that magnon and phonon modes cross each other at a point (Figure~\ref{fig:mechanism}a). In this case, even an infinitesimal coupling will lift the degeneracy, open a gap, and generate two hybrid modes. The lower mode has the capacity to carry heat and generate an odd response to a finite magnetic field. We note that  Petit \textit{et al.} \cite{Petit2021}, by performing inelastic neutron scattering, found spectroscopic evidence for mixing between phononic and magnetic modes in Tb$_3$Ga$_5$O$_{12}$, the first THE insulator~\cite{Strohm2005}.

Motivated by these considerations, we compared the computed phonon spectrum of NiPS$_3$~\cite{hashemi2017}  with the magnon spectrum extracted from inelastic neutron scattering measurements \cite{Scheie2023}. The magnetic ordering generates a new Brillouin zone  on top of the nuclear one~\cite{Wildes2022,Scheie2023}. The magnetic Brillouin projects rectangles over the hexagons of the honeycomb lattice  (See Figure~\ref{fig:mechanism}b). As a result of this zone folding, a boundary point of the nuclear Brillouin zone is located at the center of a magnetic Brillouin zone. This crucial feature allows magnons to mix up with acoustic phonons. Figure~\ref{fig:mechanism}c is a representation of  band dispersion of magnons and acoustic phonons along $\Gamma-$M.  One can see that magnon and phonon bands intersect multiply near the $\Gamma$- and M- points of the Brillouin zone. Note that these crossings, in the momentum space, occur along the $b^\ast$ and recall that in real space the amplitude of magnetization along the $b$-axis was found to set the amplitude of $\kappa_{ab}$. Near the M point, the three acoustic phonon branches cross the magnon bands and the lowest crossing occurs at energy of 3.7 meV. Finite coupling at this energy would open a gap. Yang \textit{ et al.} \cite{Yang2020}, argued that, independent of the microscopic details, an intrinsic thermal Hall effect displays a simple behavior $\kappa_{xy}/T \propto {\rm{exp}}(-T/T^\star)$. $T^\star$ represents the width of the energy window in which the Berry phase is finite and constant \cite{Yang2020}. Such a behavior has been reported in several cases \cite{Yang2020,Grissonnanche2020,Boulanger2020,Gillig2023}. Figure~\ref{fig:mechanism}d shows that our data for the least twinned sample (\#1) displays such a behavior. The fit yields $T^\star= 12$ K. This 1 meV energy scale is tantalizingly close to the distance between the magnon-phonon crossing points (Figure~\ref{fig:mechanism}c). Finally, we note that our results confirm the critical role of interlayer coupling for the persistence of antiferromagnetism in multilayer NiPS$_3~$\cite{kim2019}.

%Thus, NiPS$_3$  hosts a large $\kappa_{ab}$ whose magnitude is extremely sensitive to the anisotropy of the in-plane magnetization in contrast to the isotropic phonon mean free path. Heat-carrying acoustic phonons cross a magnon branch at an energy somewhat larger than the one extracted from the temperature dependence of $\kappa_{ab}$. Taken together, these observations provide a compelling case for an intrinsic THE in this specific case and a promising avenue for investigation of other cases. \textcolor{red}{Our observations of the thermal Hall effect and torque measurements highlight the critical role of interlayer and phonon-magnon coupling. These interactions yield significant implications for the material's properties, particularly as antiferromagnetism is lost at the monolayer limit\cite{kim2019}.

Qingkai Meng and Xiaokang Li contributed equally to this work. We thank Beno\^it Fauqu\'e, Ga\"el Grissonnanche, Roser Valenti, and particularly Andrew Wildes for helpful discussions. This work was supported by The National Key Research and Development Program of China (Grant No.2022YFA1403500), the National Science Foundation of China (Grant No. 12004123, 51861135104 and 11574097 ), and the Fundamental Research Funds for the Central Universities (Grant no. 2019kfyXMBZ071). X. L. was supported by The National Key Research and Development Program of China (Grant No.2023YFA1609600) and the National Science Foundation of China (Grant No. 12304065).\\

\noindent
* \verb|lixiaokang@hust.edu.cn|\\
* \verb|zengwei.zhu@hust.edu.cn|\\
* \verb|Liangli44@hust.edu.cn|\\
* \verb|kamran.behnia@espci.fr|\\

\bibliography{main}

\clearpage
% Add 'S' to the numbering inside the supplement
\renewcommand{\thesection}{S\arabic{section}}
\renewcommand{\thetable}{S\arabic{table}}
\renewcommand{\thefigure}{S\arabic{figure}}
\renewcommand{\theequation}{S\arabic{equation}}
\setcounter{section}{0}
\setcounter{figure}{0}
\setcounter{table}{0}
\setcounter{equation}{0}

{\large\bf Supplemental Material for ``Thermal Hall effect driven by phonon-magnon hybridization in a honeycomb antiferromagnet''}
{\large\bf by Q. Meng et al.}

\setcounter{figure}{0}

\section{Samples}
\begin{figure}[ht]
\centering
\includegraphics[width=0.9\linewidth]{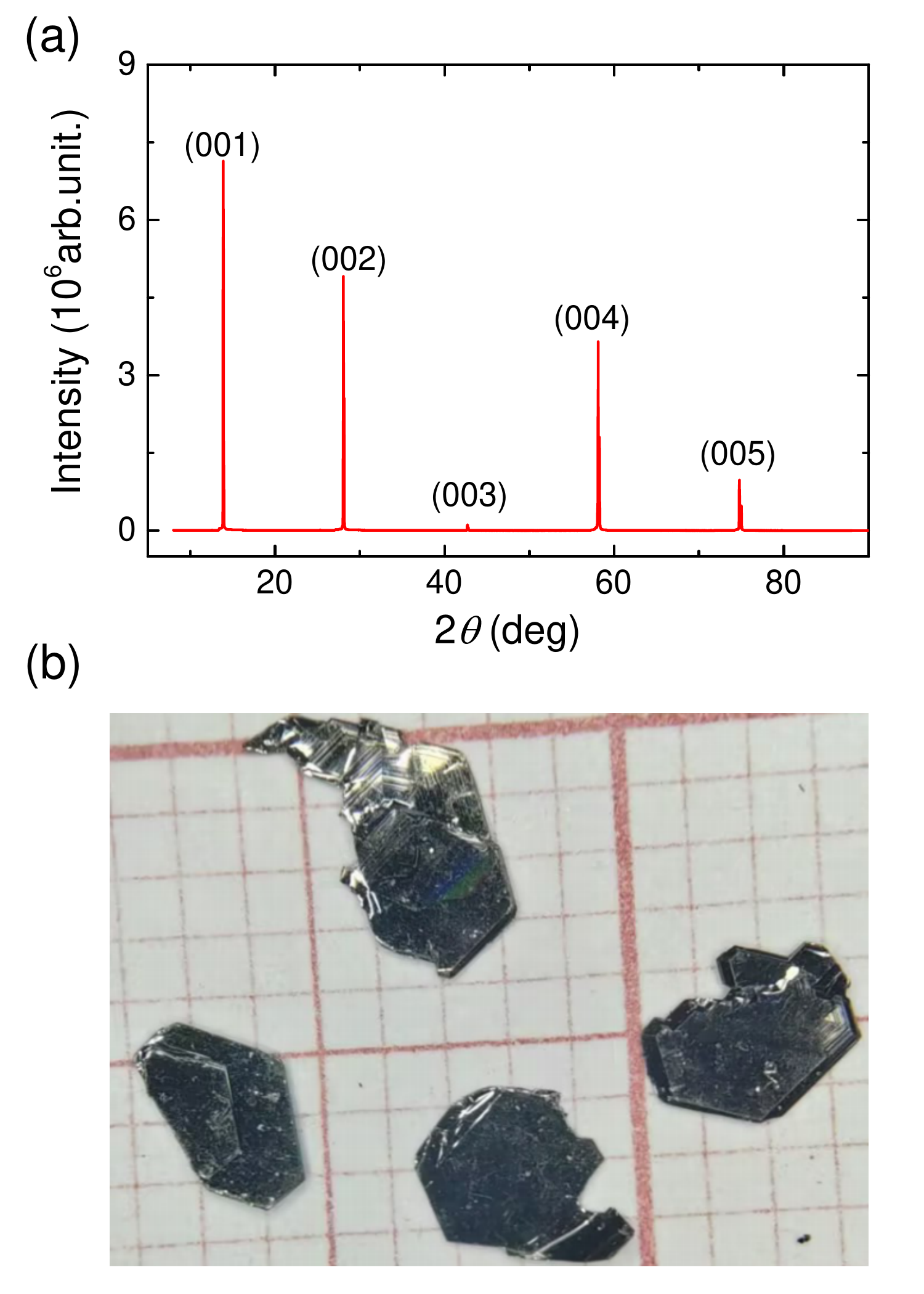} 
\caption{\textbf{X-ray and morphology of the samples} (a) X-ray powder diffraction. (b) Photograph of NiPS$_3$ single crystals.
}
\label{fig:sample}
\end{figure}

NiPS$_3$ single crystals used in this work were synthesized by the chemical vapor transport(CVT) method using I$_2$ as the transport agent. Polycrystalline NiPS$_3$ was first synthesized via the solid state reaction of high purity elements of Ni (powder, Alfa, 99.996\%), P (lump, Alfa, 99.999\%) and S (pieces, Alfa, 99.999\%). The mixture were weighed in a stoichiometric ratio and sealed in an evacuated quartz tube and subsequently heated up to 750  $^\circ $C for 50 hours. Then, the obtained materials were ground, sealed in an evacuated quartz tube with iodine (powder, Alfa, 99.99\%) concentration of 3 mg/cm$^3$. The tube was transferred into a two-zone furnace with a temperature gradient from 750 $^\circ $C to 700  $^\circ $C for one week. The samples show strong diffraction peaks and the hexagonal shape with the size up to millimeters, as shown in Figure~\ref{fig:sample}. They were cut into rectangles (with the long and short side along $a$ and $b$ axes respectively) and the required dimensions by a wire saw, to fit the thermal transport measurements.

\begin{figure}[ht]
\centering
\includegraphics[width=0.90\linewidth]{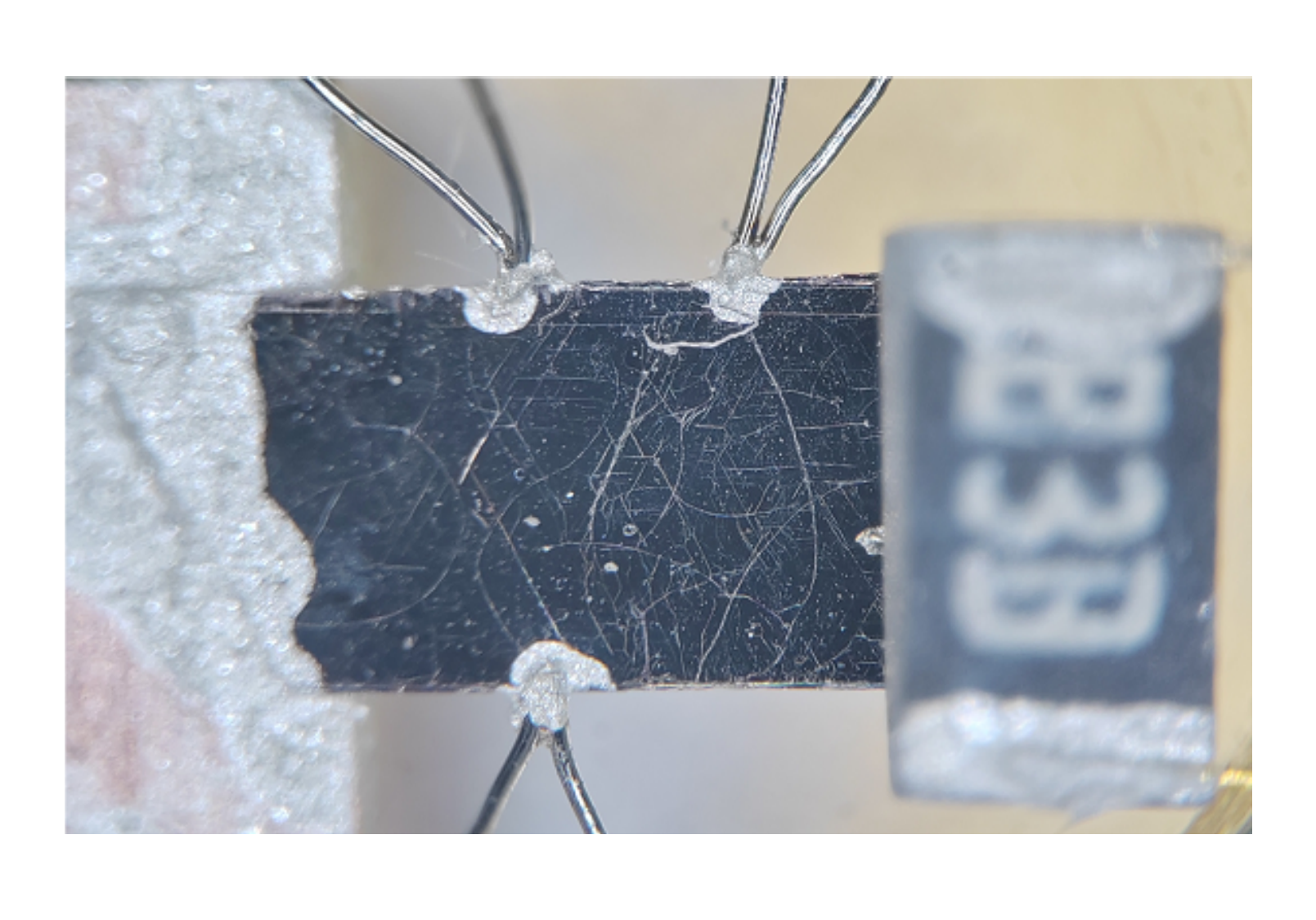} 
\caption{\textbf{Optical picture of the sample mounted on the heat sink.}
}
\label{fig:optica-picture}
\end{figure}

\begin{figure}[ht]
\centering
\includegraphics[width=1.0\linewidth]{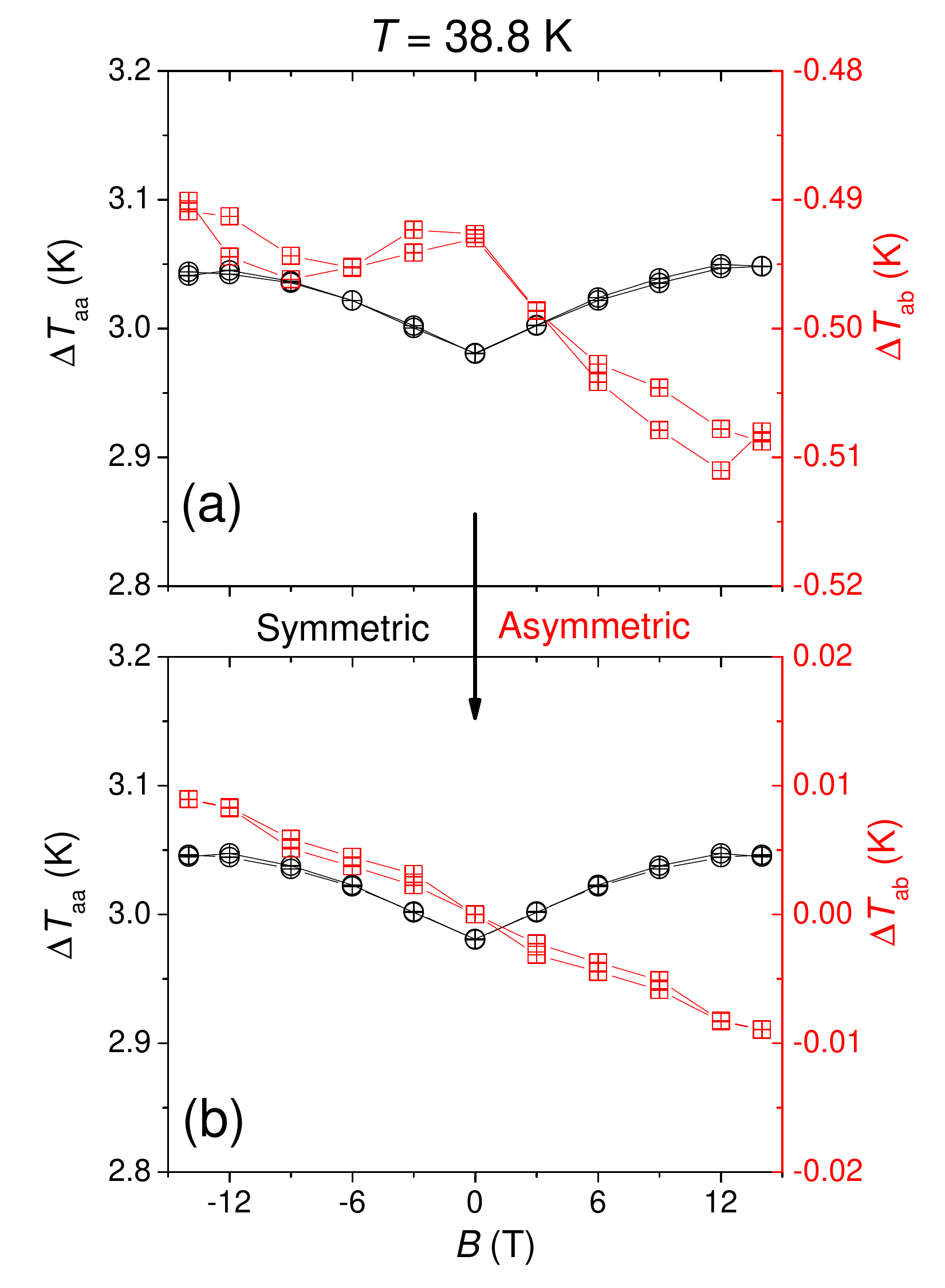} 
\caption{\textbf{Raw data before and after extracting the even (symmetric) and the odd (anti-symmetric) responses.}
}
\label{fig:raw-data}

\end{figure}
\begin{figure}[ht]
\centering
\includegraphics[width=1.0\linewidth]{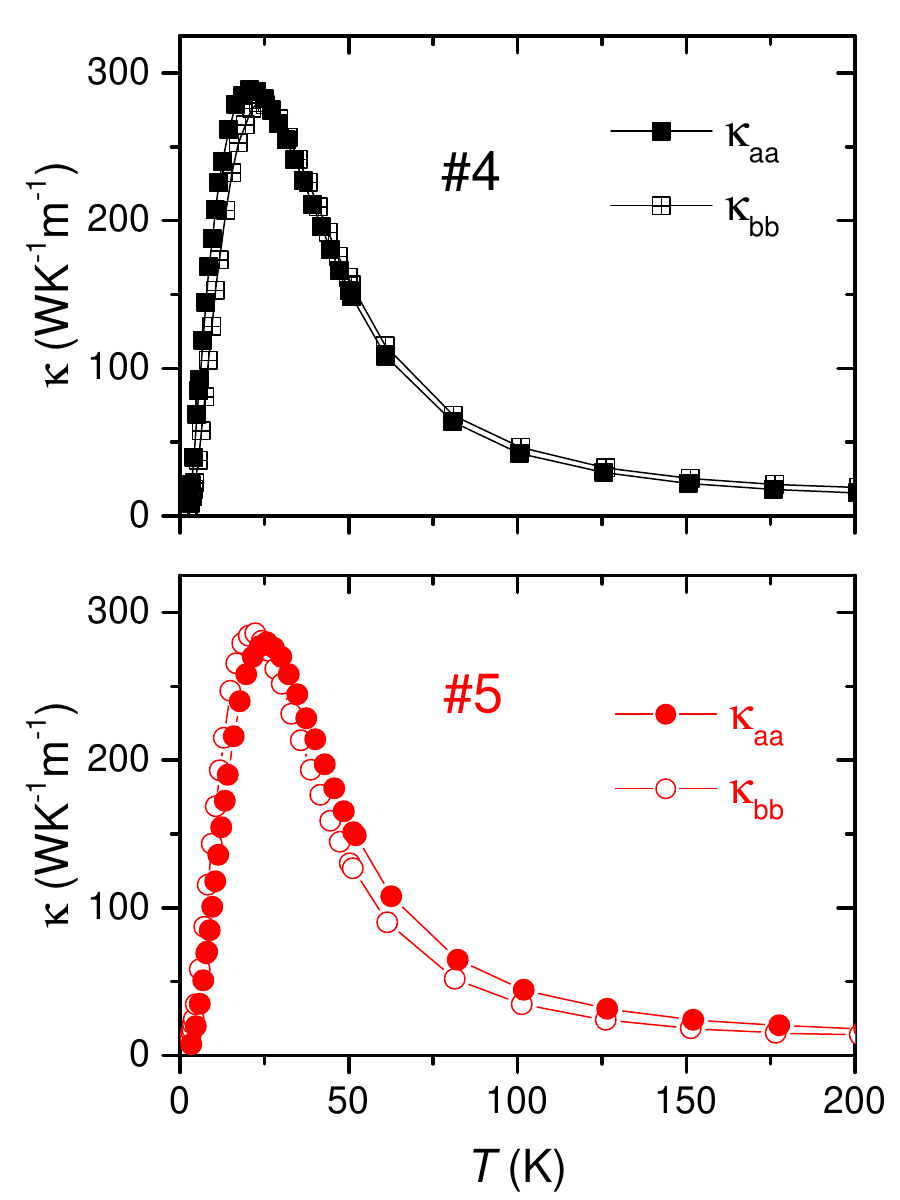} 
\caption{\textbf{Isotropic in-plane longitudinal thermal conductivity.} Temperature dependence of $\kappa_{aa}$ and $\kappa_{bb}$ in two NiPS$_{3}$ samples.}

\label{fig:Kappa_ii}
\end{figure}

\begin{figure}[ht]
\centering
\includegraphics[width=1.0\linewidth]{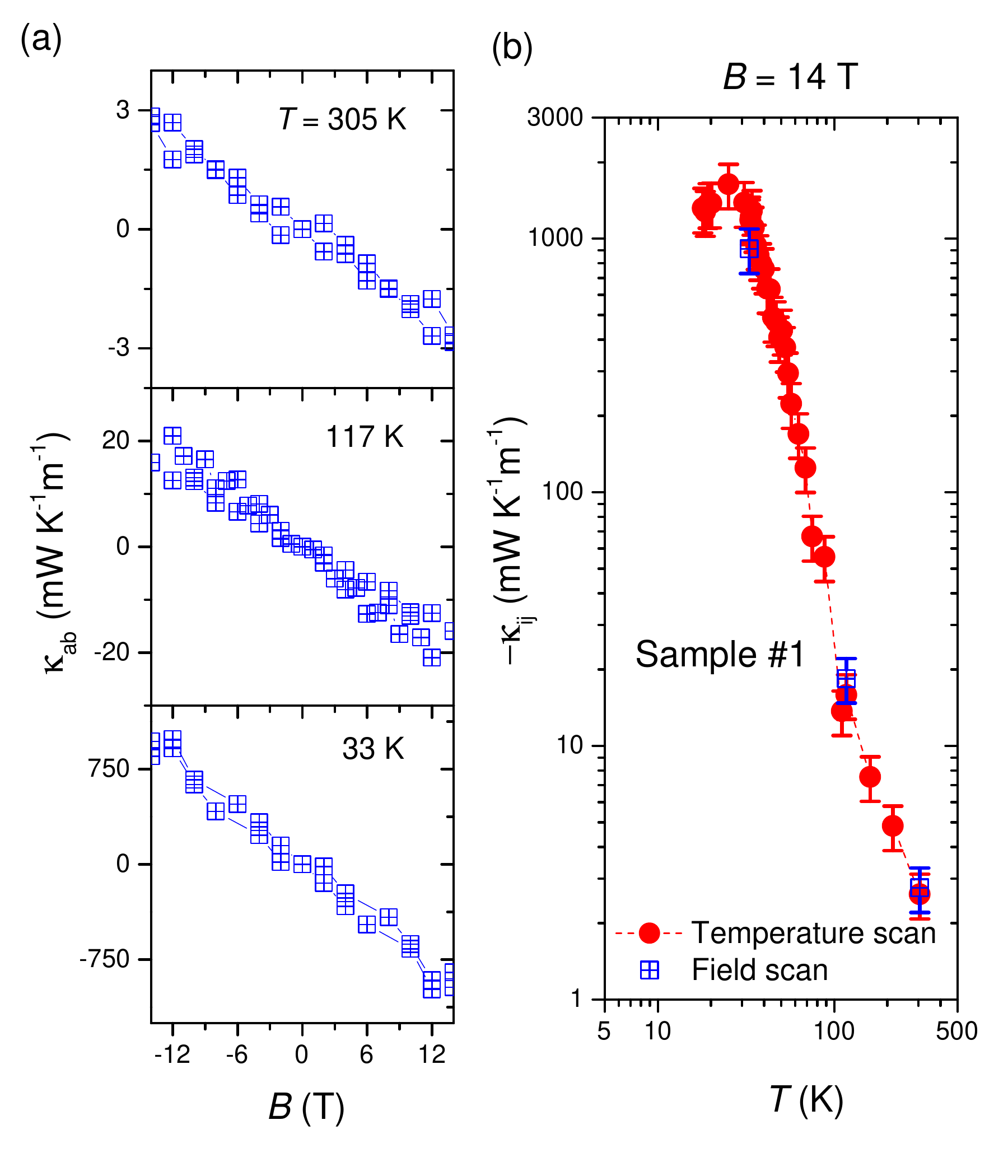} 
\caption{\textbf{Comparison of field-scan and temperature-scan thermal Hall data.} The results  match with each other.
}
\label{fig:scan-methods}
\end{figure}

\section{Measurements}
All thermal transport and specific heat experiments were performed in a commercial physical property measurement system (Quantum Design PPMS) within a stable high-vacuum sample chamber. The one-heater-three-thermocouples (type E) method was used for measuring the longitudinal and transverse thermal gradient simultaneously. The thermal gradient in the sample was produced through a 7 k$\Omega$ chip resistor alimented by a current source (Keithley6221). The DC voltage on the heater and thermocouples was measured through the DC-nanovoltmeter (Keithley2182A). The thermocouples, the heat-sink, and the heater were connected to samples directly by the silver paste  (see Figure~\ref{fig:optica-picture}). All thermal transport measurements were performed on the same setup, that was checked and calibrated by measuring the glass. Figure~\ref{fig:raw-data} is the raw data to show the symmetric and asymmetric processes for the longitudinal and transverse temperature difference. The longitudinal ($\nabla T_x = -(T_3-T_2)/l$) and the transverse ($\nabla T_y = -(T_2-T_1)/w$ ) thermal gradients generated by a longitudinal thermal current $J_Q$ lead to the longitudinal ($\kappa_{aa}$) and the transverse ($\kappa_{ab}$) thermal conductivity by the following formulas:
\begin{equation}\label{kappaii}
\kappa_{aa} = \frac{Q}{\nabla T_a}
\end{equation}
\begin{equation}\label{kappaij}
\kappa_{ab} = \frac{\nabla T_b}{\nabla T_a} \cdot \kappa_{aa}
\end{equation}
Here $l$, $w$, $Q$ are the distance between longitudinal thermocouples, the sample width and the heat power respectively.   The isotropic in-plane thermal conductivity $\kappa_{aa} = \kappa_{bb}$ is assumed and verified in two different NiPS$_3$ samples (see Figure \ref{fig:Kappa_ii}). Both field-scan and temperature-scan methods were used for the thermal Hall measurements. The results match well, as seen in Figure~\ref{fig:scan-methods}.

Magnetization below 7 T were measured in a commercial Superconducting Quantum Interference Device (Quantum Design SQUID). Field-dependent magnetization up to 55 T were measured in a pulsed magnetic field equipment at the Wuhan National High Magnetic Field Center (WHMFC).

\section{Specific heat} 
\begin{figure}[ht]
\centering
\includegraphics[width=0.95\linewidth]{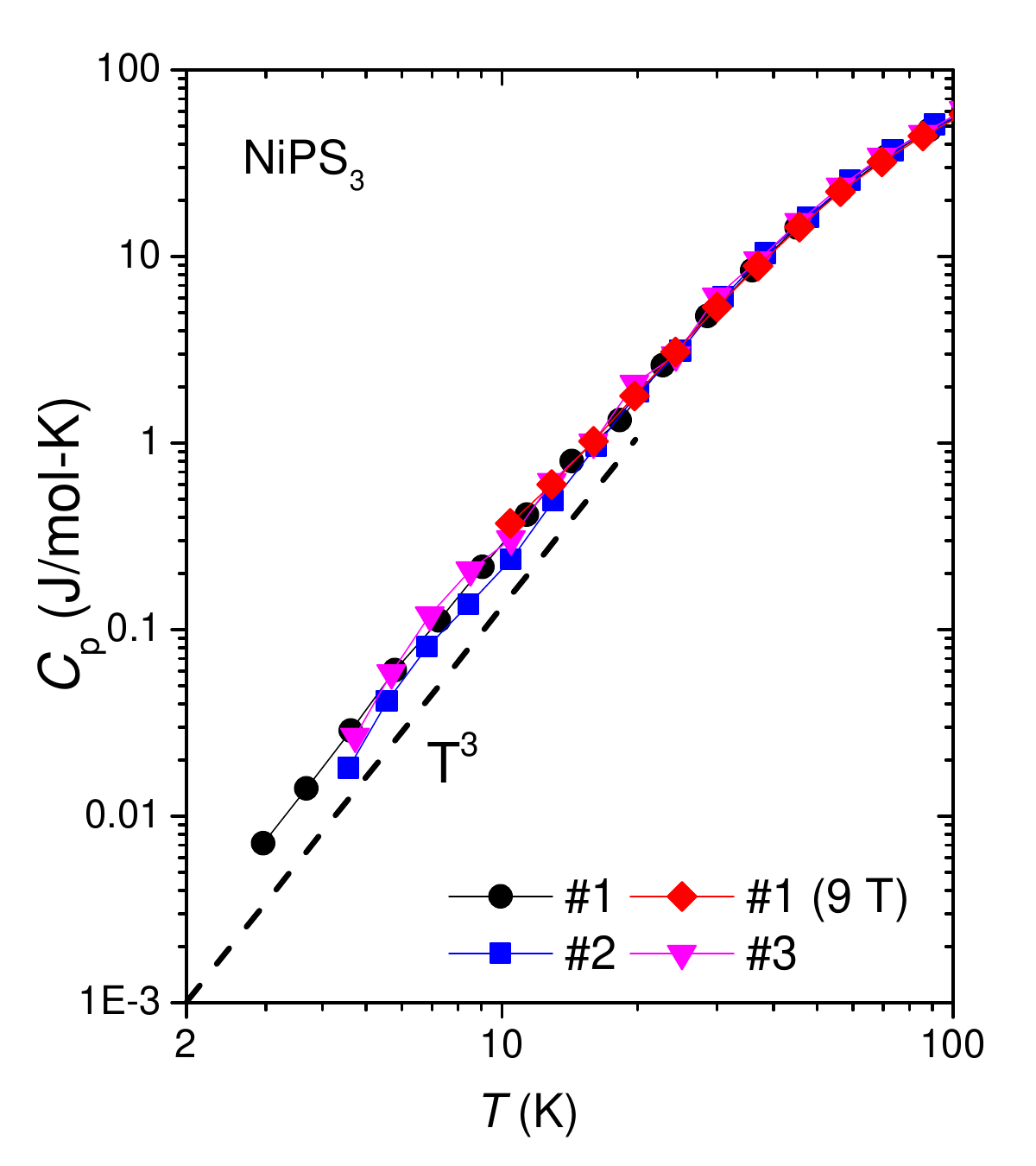} 
\caption{\textbf{specific heat.} Temperature dependence of  specific heat in different NiPS$_{3}$ samples at zero and finite magnetic field. 
}
\label{fig:specific heat}
\end{figure}

Figure~\ref{fig:specific heat} compares specific heat in samples $\#1$, $\#2$ and $\#3$ at zero field and at 9 T. All curves exhibit a coincident behavior, meaning that the mean free path $\ell = 3\kappa/C\upsilon$ is the same in different samples, and almost unchanged by field. The prefactor of the cubic temperature dependence of the specific heat yields a Debye temperature of $\Theta_D= 210$ K.

\begin{figure}[ht]
\centering
\includegraphics[width=0.8\linewidth]{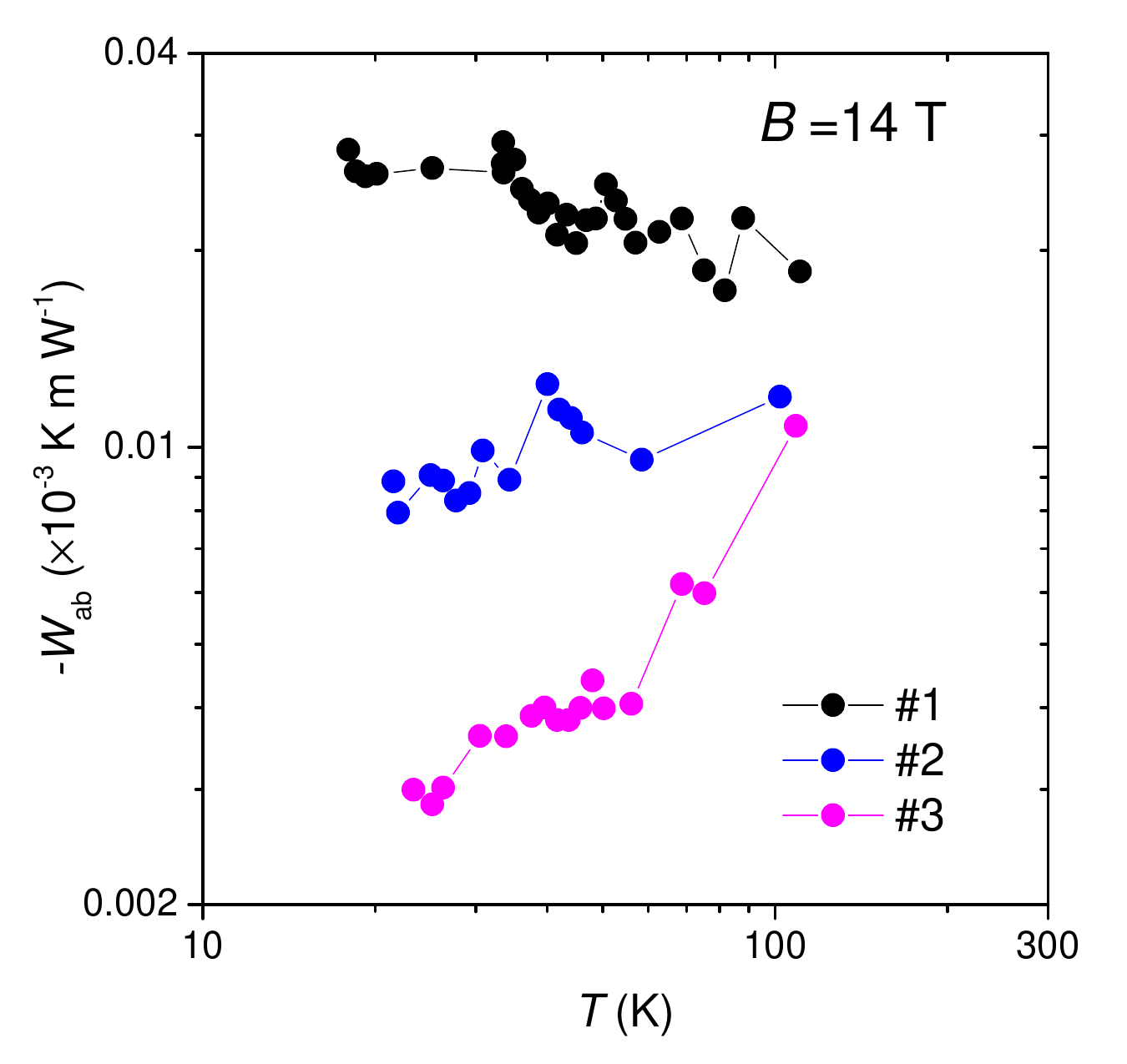} 
\caption{\textbf{Thermal Hall resistivity in different samples.} The results is extracted from the data in Fig.3d and e in the main text by the formula $\kappa_{ab}$ = $w_{ab} \cdot \kappa_{aa}^2$. 
}
\label{fig:thermal-Hall-resistivity}
\end{figure}

\begin{figure}[ht]
\centering
\includegraphics[width=1\linewidth]{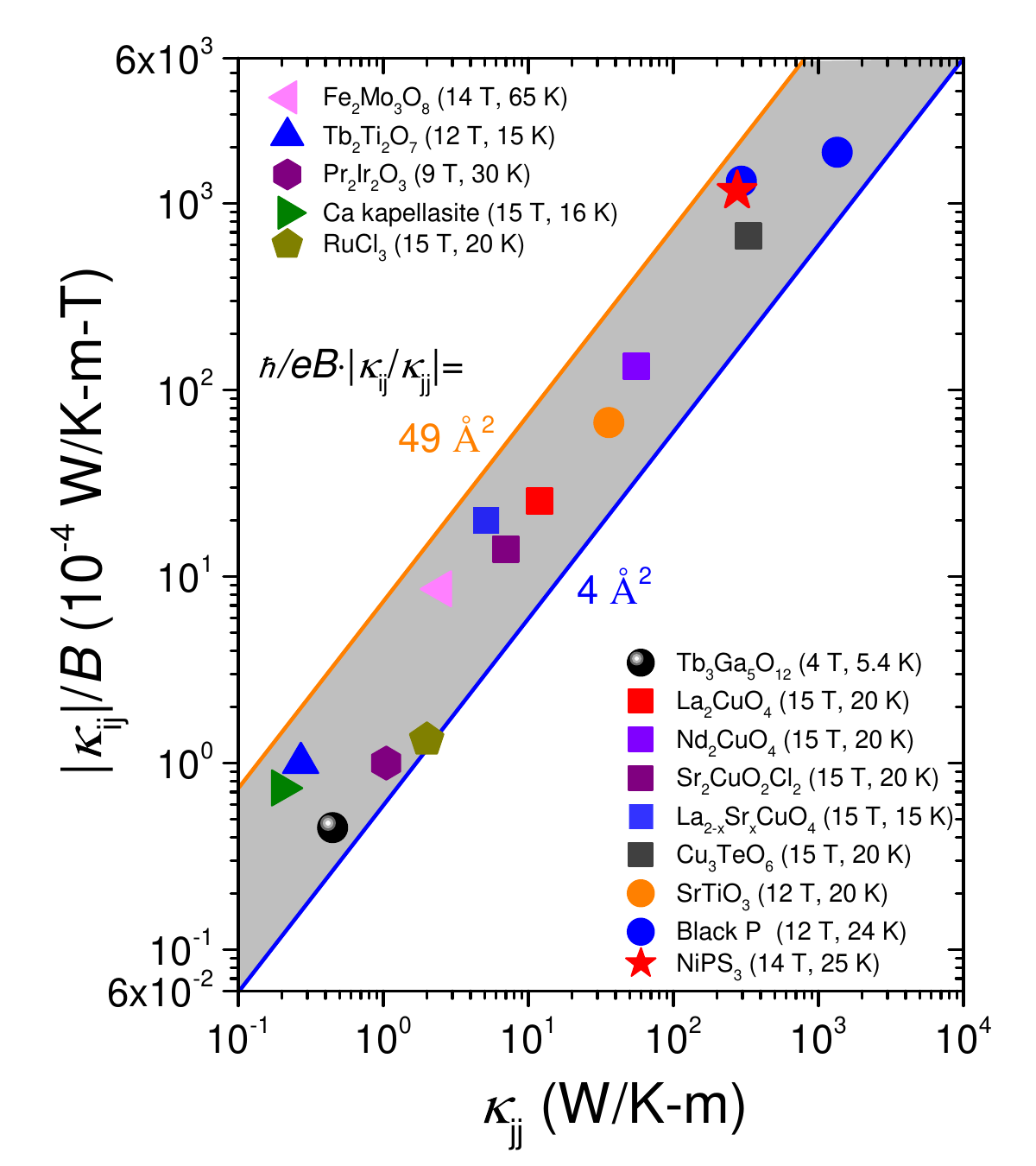} 
\caption{\textbf{Thermal Hall angle in different insulators.} The transverse thermal conductivity divided by magnetic field as a function of longitudinal thermal conductivity in different insulators (source: ~\cite{Strohm2005, Ideue2017, Bruin2022, Grissonnanche2019, Grissonnanche2020, Boulanger2020, Li2020, Uehara2022, Chen2022, Doki2018, Li2023}). $\kappa_{ij}$ takes its peak value and $\kappa_{jj}$ takes its value at the corresponding temperature. Only cases with a linearly or sub-linear field dependence are included.}
\label{fig:comparision}
\end{figure}

\section{Thermal Hall resistivity}
Figure \ref{fig:thermal-Hall-resistivity} shows thermal Hall resistivity in different samples, which is extracted from the data in Fig.3d and e in the main text by the formula $\kappa_{ab}$ = $w_{ab} \cdot \kappa_{aa}^2$. It can  be clearly demonstrated that the spread of these curves in Fig.3e does not arise from squaring a small difference of $\kappa_{aa}$ in Fig.3d.

\section{The thermal Hall angle in different insulators} 
Figure~\ref{fig:comparision} compares longitudinal ($\kappa_{jj}$) and transverse ($\kappa_{ij}/B$) thermal conductivities of different insulators. Both the thermal Hall conductivity and the thermal Hall angle of NiPS$_3$ are close to the record among insulators. In different insulators, the longitudinal thermal conductivity $\kappa_{jj}$ varies by 4 orders of magnitude, but the $\kappa_{ij}/\kappa_{jj}/B$ ratio remains within the range of $\approx 10^{-4}$-$10^{-3}$ T$^{-1}$, indicating a length scale $\lambda_{tha}=\ell_B \cdot\sqrt{\kappa_{ij}/\kappa_{jj}}$ remains between 2 and 7 \AA, comparable to the shortest phonon wavelength allowed by the distance between atoms. 

%\textcolor{red}{For magnonic thermal Hall effect in some magnetic insultors, such as Lu$_2$V$_2$O$_7$~\cite{Onose2010} and Cu(1,3-bdc)~\cite{Watanabe2016}, the former with a spontaneous signal of $\kappa_{ij}/\kappa_{jj}$  = $2\cdot 10^{-3}$ is hard to be compared in Figure~\ref{fig:comparision}, the lattar with a peak signal of $\kappa_{ij}/\kappa_{jj}/B$ = 1.5$\cdot$ $10^{-5}$ T$^{-1}$ is far below the lower limit of the narrow range in Figure~\ref{fig:comparision}.}

\section{Comparison with Kitaev candidates}
The Kitaev model \cite{KITAEV20062} describing a honeycomb lattice of spins with bond-dependent interactions has attracted much recent attention because it is an exactly solvable model of a spin liquid. The compound $\alpha$-RuCl$_3$ has emerged as a prominent candidate for Kitaev physics \cite{Plumb2014}. It has a zigzag antiferromagnetic ground state with a N\'eel temperature of $\approx$ 7 K, which can be destroyed by a magnetic of $\approx$ 10 T. The observation of a large thermal Hall effect in  $\alpha$-RuCl$_3$ \cite{Kasahara2018,Hentrich2019,Yamashita2020,Lefran2022,Bruin2022} has attracted much recent attention. Another Kitaev candidate in which a thermal Hall signal has been detected is Na$_2$Co$_2$TeO$_6$ \cite{Gillig2023}.

\begin{figure*}[ht]
\centering
\includegraphics[width=0.8\linewidth]{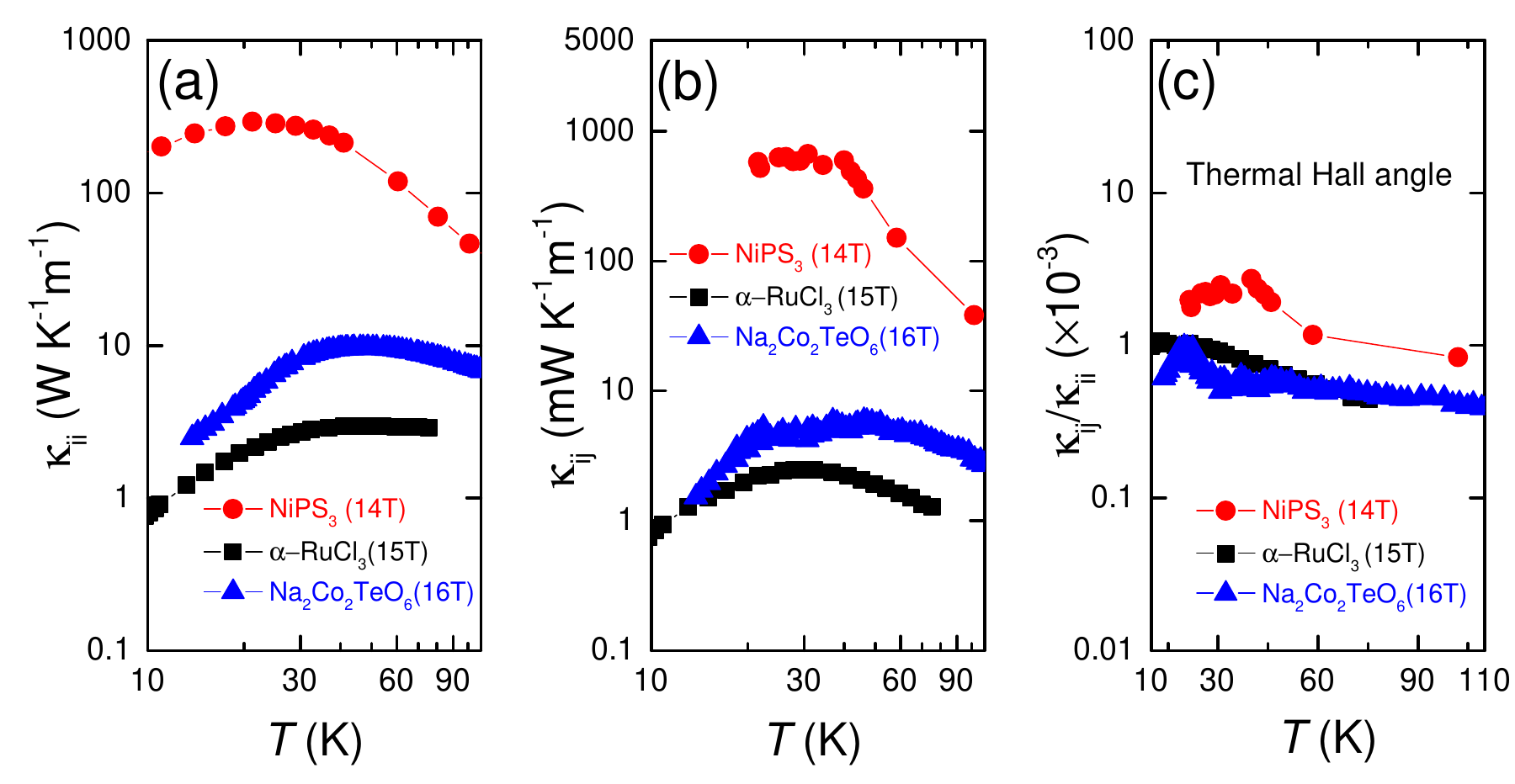} 
\caption{\textbf{Comparison with two Kitaev candidates.} (a)-(b) Longitudinal and transverse thermal conductivity in NiPS$_3$, $\alpha$-RuCl$_3$ ~\cite{Lefran2022} and Na$_2$Co$_2$TeO$_6$ \cite{Gillig2023}. (c) Temperature dependence of the thermal Hall angle. Note that the panel a and b were shown in log-log plot, and the panel c was shown in semi-log plot.}
\label{fig:NiPS and RuCl}
\end{figure*}

Figure \ref{fig:NiPS and RuCl} compares the longitudinal and the transverse thermal conductivity in NiPS$_3$($B =$ 14 T), in $\alpha$-RuCl$_3$($B$ = 15 T),  and in Na$_2$Co$_2$TeO$_6$ ($B=$16 T). Both are two orders of magnitude larger in NiPS$_3$ compared to the other two. The amplitude of $\kappa_{xy}$ reported by different groups \cite{Kasahara2018,Hentrich2019,Yamashita2020,Lefran2022,Bruin2022} is not identical. This difference has been attributed to the details of sample growth \cite{Yamashita2020,Bruin2022}. However, this sample dependence remains orders of magnitude smaller than the 100-fold variation between NiPS$_3$  and $\alpha$-RuCl$_3$. In other words, while the amplitude of maximum phonon mean free path in different $\alpha$-RuCl$_3$ samples are somewhat different, they remain all $\approx$ 100 times shorter than the phonon mean free path in twinned and untwinned NiPS$_3$ crystals. It remains to be seen if this shortening of the phonon mean free path reflects a difference in the nature of magnetic excitations in the two cases.

On the other hand, the thermal Hall angle($\kappa_{ij}$/$\kappa_{ii}$) in the three solids peaks to the same order of magnitude (Figure \ref{fig:NiPS and RuCl})(c), implying  that the  mean free path of phonons does not affect the thermal Hall angle. 

\section{The exponential temperature dependence of $\kappa_{ab}/BT$}
\begin{figure*}[ht]
\centering
\includegraphics[width=0.8\linewidth]{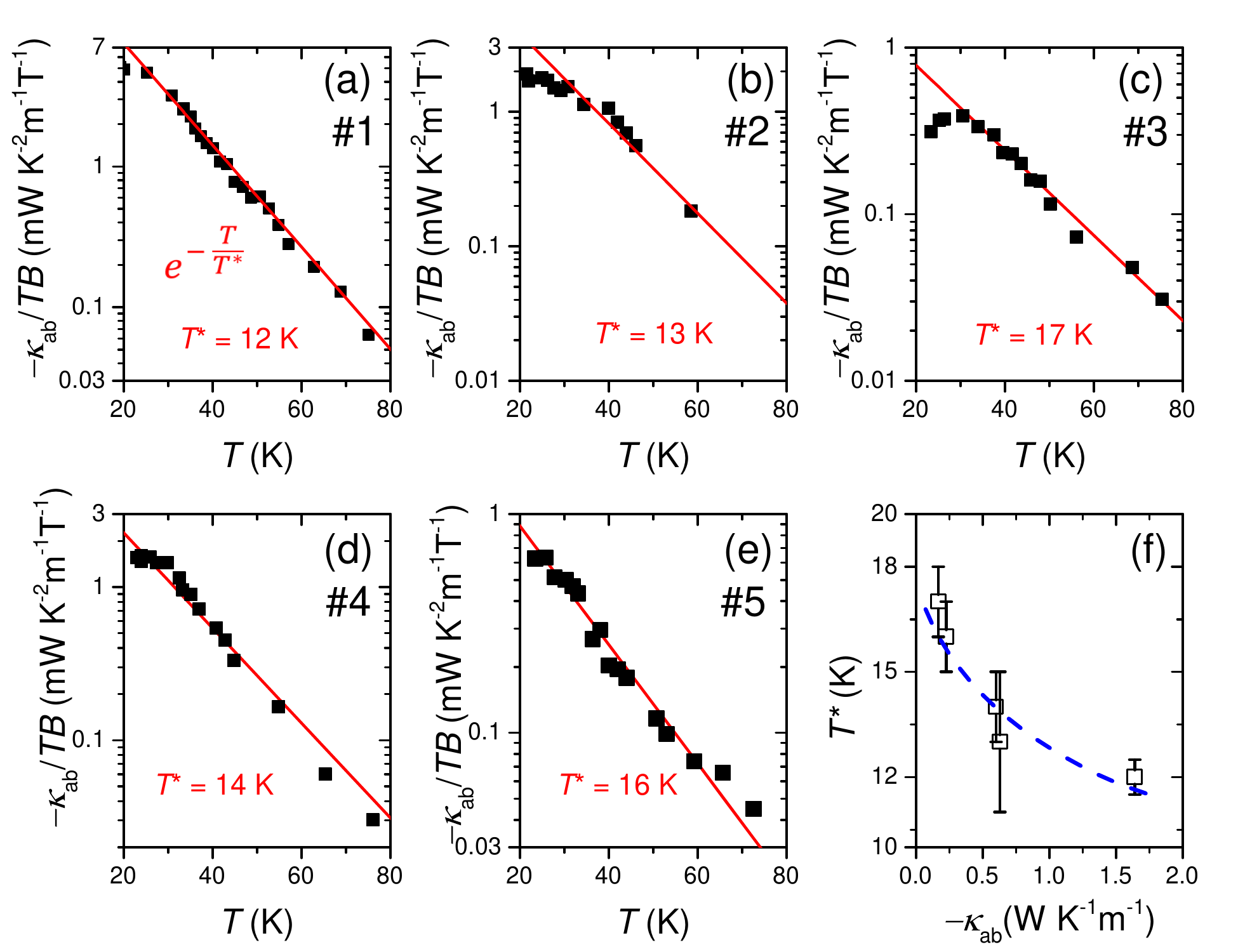} 
\caption{\textbf{Exponential fitting in five different samples.} (a-e) Fitting the $\kappa_{ab}/TB$ using the exponential formula of $e^{-T/T^*}$ in five different samples and in the intermediate temperature region. (f) Comparison of the characteristic temperature $T^*$ extracted by the fit and the amplitude of $\kappa_{ab}$. The blue dashed line is a guide for the eye. }

\label{fig:fitting kappa over B}
\end{figure*}

% \textcolor{red}{Figure \ref{fig:fitting kappa over B} shows the exponential fitting and polynomial fitting of $\kappa_{ab}/B$. The comparison shows that the exponential fitting is more consistent with the experimental data. } 
Figure \ref{fig:fitting kappa over B} (a)-(e) shows that the exponential fitting in five different samples in the intermediate temperature region. The characteristic temperature $T^*$ extracted by fitting may reflect the energy gap formed by phonon-magnon hybridization, which is correlated with the amplitude of $\kappa_{ab}$, as seen in {Figure \ref{fig:fitting kappa over B} (f).}

%\begin{figure*}[ht]
%\centering
%\includegraphics[width=0.9\linewidth]{SM-12-comparison-different-fitting.pdf} 
%\caption{
%\textbf{Comparison of exponential fitting and polynomial fitting result of $\kappa_{ab}/B$.} Note that the panel a and b were shown in semi-log and log-log plot respectively.}
%\label{fig:exp fitting}
%\end{figure*}

\begin{figure*}[ht]
\centering
\includegraphics[width=1.00\linewidth]{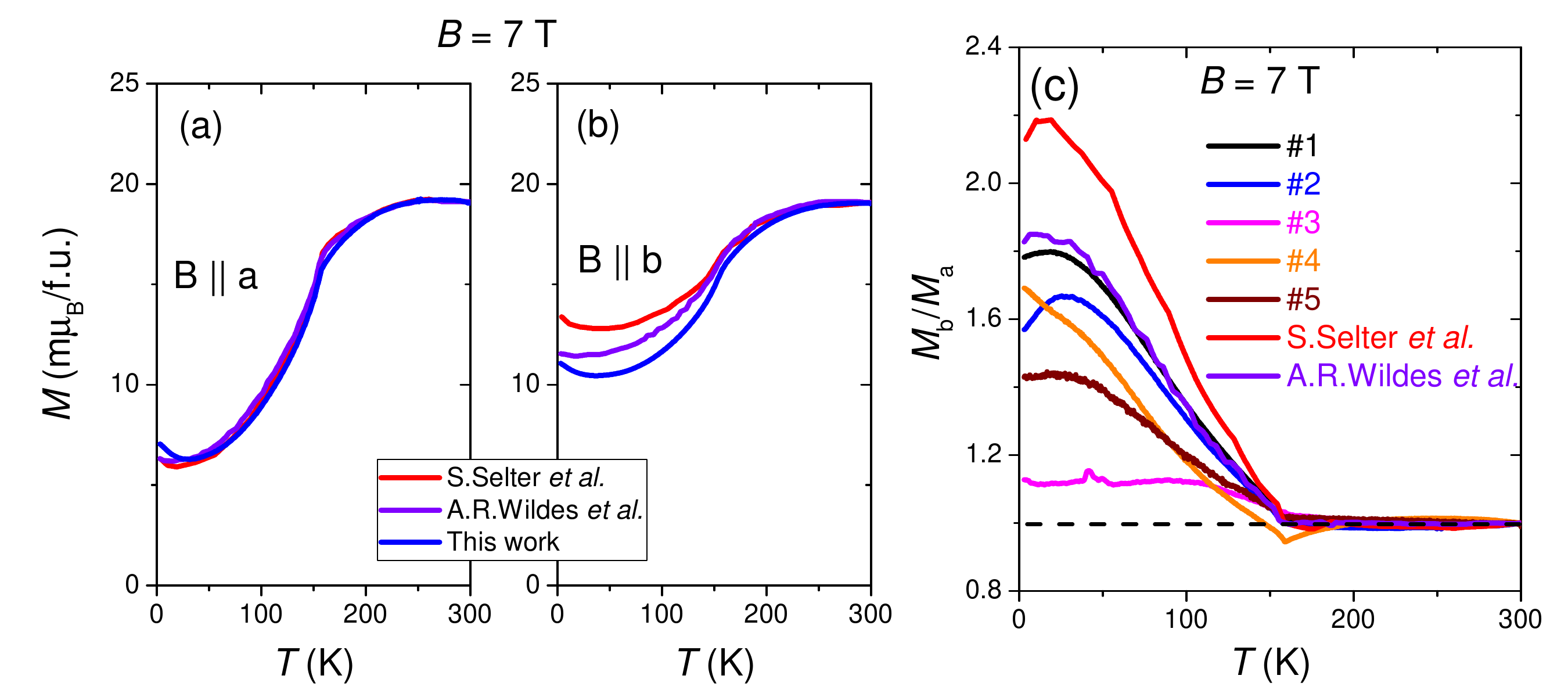} 
\caption{\textbf{In-plane magnetic anisotropy} (a-b) Comparison of $M_a$ and $M_b$ reported by different groups\cite{Wildes2015,Selter2021}. (c) Temperature dependent in-plane magnetic anisotropy ($M_b/M_a$) in different NiPS$_3$ crystals. }
\label{fig:Magnetic anisotropy}
\end{figure*}

\section{In-plane magnetic anisotropy} 
Figure \ref{fig:Magnetic anisotropy}a and b shows the comparison of $M_a$ and $M_b$ reported by different groups. $M_a$ is virtually identical, $M_b$ displays sample dependence. Figure \ref{fig:Magnetic anisotropy}c shows the in-plane magnetic anisotropy ($M_b/M_a$) as a function of temperature. It has the same sample trend with the $\kappa_{ab}$, largest in sample $\# 1$ and smallest in sample $\# 3$.

\section{Twins and  magnetic anisotropy} 
\begin{figure*}[ht]
\centering
\includegraphics[width=0.8\linewidth]{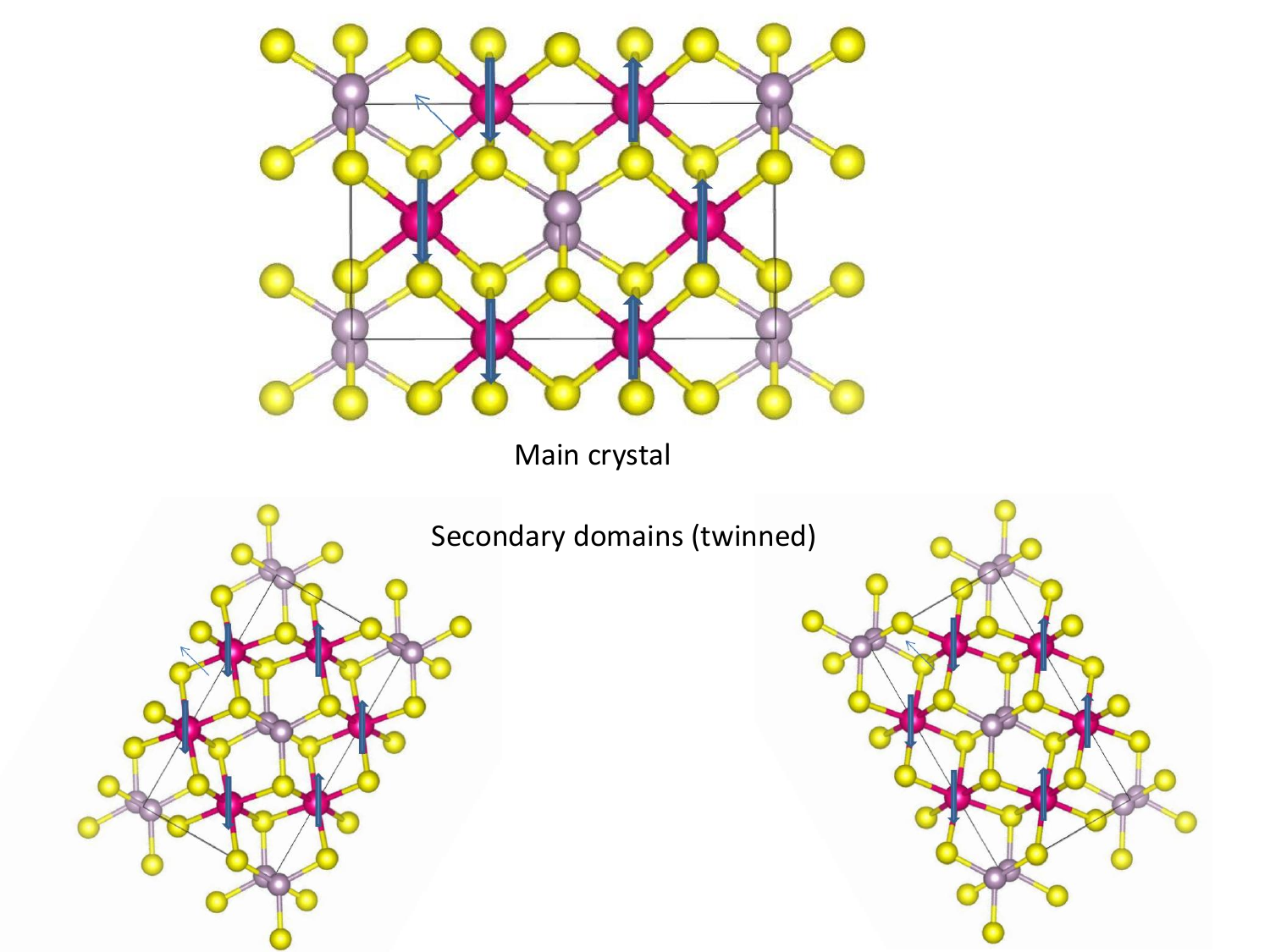} 
\caption{\textbf{Twins.} Schematic diagram of spins and atomic bonds in an untwinned crystal top and in two minority domains generated by twininng. The angle between spins and atomic bonds are different. This attenuates magnetic anisotropy.
}
\label{fig:Twinned}
\end{figure*}

%We found that NiPS$_3$ crystals display a remarkable correlation between the amplitude of the thermal Hall conductivity and the in-plane magnetic anisotropy and attributed the variation and correlation to the presence of twins. In an untwinned crystal (see Figure~\ref{fig:Twinned}a), in-plane magnetic anisotropy is large. When the stacked structure  is twinned (see Figure~\ref{fig:Twinned}b), the in-plane magnetic anisotropy is attenuated. This would explain the difference between sample \#1 and sample \#3. When the temperature decreases below $T_N$, the $a$- and $b$ lattice parameters decrease and increase respectively (see in Figure \ref{fig:Twinned}c)\cite{murayama2016}). In a twinned crystal, there would be an alternation of shrinking $a$-axis and expanding $b$-axis. As a result, $M_a$ will mix up with $M_b$ and $\kappa_{ab}$ will mix up with  $\kappa_{ba}$. As a result, in a twinned crystal with equal population of three domains, One would find $\kappa_{ab}=0$ and  ($M_b$-$M_a$) = 0. This explains the correlation seen in Figure \ref{fig:sampledependent}d. 
A study employing transmission electron microscopy and powder X-ray diffraction  \cite{murayama2016} concluded that FePS$_3$ forms a rotational twin structure with the common axis along the $c^\star$-axis. The twin boundaries were found to be at the van der Waals gaps between the layers (See Fig. \ref{fig:Twinned}).  A more recent inelastic neutron scattering \cite{Scheie2023} found similar twins in NiPS$_3$. In the latter study, X-ray Laue diffraction could not distinguish between the [100], the [-1/2,1/2,0] and the [-1/2,-1/2,0] directions. This mean that the sample was a stack of layers in which the true $a$-axis and $b$-axis were off 120 degrees each other.

The spin lattice has a zigzag structure with spins oriented along the $a$-axis with a small canting angle. The crystal structure has an approximate threefold symmetry. Therefore in presence of twins, the angle between the spins and the atomic bonds is different in the three crystalline domains. As a consequence, the presence of twins  (Fig.\ref{fig:Twinned}) attenuates the in-plane magnetic anisotropy leading to the sample dependence seen in Figure \ref{fig:Magnetic anisotropy}.

\begin{figure*}[ht]
\centering
\includegraphics[width=0.7\linewidth]{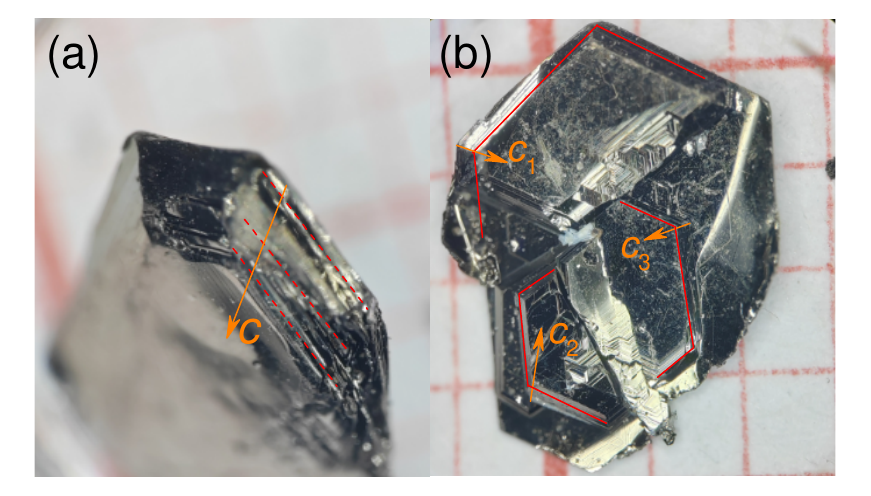} 
\caption{\textbf{Visualising twins in NiPS$_3$.}  (a) The picture of a twinned crystal. (b) The same crystal turned over reveals three distinct $c$-axes noted as $c_1$, $c_2$ and $c_3$.
}
\label{fig:photo of twin}
\end{figure*}

Figure \ref{fig:photo of twin} shows a photograph of one of our NiPS$_3$ crystals showing visible twinning along the  $c^\star$-axis. In Figure \ref{fig:photo of twin}(a), one clearly sees a stacking feature along the $c$ direction. In Figure \ref{fig:photo of twin}(b), the red lines mark three twinned crystals with different stacking orientations, respectively $c_1$, $c_2$ and $c_3$.

%\textcolor{red}{\section{Twinning and mixing}}
%\textcolor{red}{Figure \ref{fig:Twinnng-mixing} shows the comparison of twinning and mixing. In twinning, the interlayer rotation angle is 120 degrees, and the in-plane lattice axis and spin easy axis doesn't change. The sample is still a single crystal with a perfect surface appearance, and will show an obvious in-plane magnetic anisotropy, as shown in Figure \ref{fig:Twinnng-mixing} (a). In mixing, the interlayer rotation angle is arbitrary, there is no strict in-plane lattice axis and spin easy axis anymore. The sample presents a near-polycrystalline appearance, and the magnetic anisotropy will be weakened, as shown in Figure \ref{fig:Twinnng-mixing} (b).}

%\begin{figure*}[ht]
%\centering
%\includegraphics[width=0.9\linewidth]{SM-15-twinning-and mixing-v1.pdf} 
%\caption{\textcolor{red}{\textbf{Comparison of twinning and mixing.} (a) The interlayer rotation angle of the twinning is 120 degrees, and the in-plane lattice axis and spin easy axis have not changed. The sample is still a single crystal with a perfect surface appearance, and will show an obvious in-plane magnetic anisotropy. (b) The interlayer rotation angle of mixing is arbitrary, there is no strict in-plane lattice axis and spin easy axis anymore. The sample presents a near-polycrystalline appearance, and the magnetic anisotropy will be weakened. 
%}}
%\label{fig:Twinnng-mixing}
%\end{figure*}

\end{document}